\documentstyle[12pt]{article}
\begin{document}

\title{Constrained Quantization of Charged Strings in Background $B$ Field and $g$-Factors}

\author{
  Akira Kokado\thanks{E-mail: kokado@kobe-kiu.ac.jp},\  
  Gaku Konisi\thanks{E-mail: konisi@kgupyr.kwansei.ac.jp} \  
  and \ 
  Takesi Saito\thanks{E-mail: tsaito@yukawa.kyoto-u.ac.jp} \\
  \\
 {\small{\it{Kobe International University, Kobe 655-0004, Japan${}^{\ast }$}}} \\
 {\small{\it{Department of Physics, Kwansei Gakuin University, Nishinomiya 662-8501, 
 Japan${}^{\dagger ,\ddagger }$}}}}

\date{\small{Sept. 2000}}
\maketitle

\begin{abstract}
 The Dirac quantization is performed for the constrained system of the open string with different charges located at both ends in the constant background $B$ field. Noncommutativity reveals to commutators $[X, X]$, $[P, P]$ and also $[X,P]$ at both ends of the string. We consider a dependence on the change of the "cyclotron frequency" of the charged string. The $g$-factor of the charged string is also calculated in the framework of our formulation. 
\end{abstract}

\setlength{\parindent}{1cm}
\newpage

\renewcommand{\thesection}{\Roman{section}.}
\renewcommand{\theequation}{\arabic{section}.\arabic{equation}}
\setcounter{equation}{0}

\section{Introduction}
\indent

 The idea of noncommutative structure at small length scales is not new. In 1947 Snyder first considered "quantized space-time."  The idea was that noncommutative spacetimes could introduce an effective cut-off in field theory similar to a lattice. Recently there has been a revival of this idea and published many papers[1]. Especially interesting is the model of open strings propagating in a constant two-form ($B$ field) background. Previous studies show that this model is related to noncommutativity of $D$-branes[2,3], and in the zero slope limit to noncommutative Yang-Mills theory[4]. The equivalence of the ordinary gauge fields and the noncommutative gauge fields has then been proposed[5]. Recently the equivalence has been discussed in string field theories based on the noncommutative algebra[6]. An intriguing mixing of the UV and IR has also been found in the perturbative dynamics of noncommutative field theories[7]. \\ 
\indent     
 Canonical quantization of the open string in the presence of D-branes with a constant $B$ field has so far been discussed in the literature[8], treating the mixed boundary conditions as primary constraints and employing the Dirac quantization method. This is, however, restricted to quantization of the neutral string with opposite charges at both ends. In the present paper we would like to generalize this quantization to the case of the open string with different charges at both ends (charged string) in the $B$ field. The spectrum of charged open strings in the $B$ field was first considered in Ref.[9], and recently in Ref.[10]. In these works, however, the Dirac quantization had not yet been considered. This generalization is important, though it is rather exhausting. \\
\indent 
 In Secs. II-V the Dirac brackets are calculated. In Sec. VI we find noncommutativity for $X$ and $X$, and also for $P$ and $P$ at both ends of the charged string, where $P$ is the canonical momentum of $X$. We also find that there is a noncommutative term in the commutator $[X,P]$ except for the delta function. In the remaining sections we consider several topics: VII. Mode expansions, VIII. Virasoro algebra, IX. Dependence on the "cyclotron frequency",
X. The $g$-factors of charged strings, XI. Conclusions. \\
\indent
 The $g$-factors of charged strings have already been caluculated in the old ordinary formulation[11]. Here, we would like to calculate them again more completely in the framework of our formulation, which leads us to the noncommutative strings.

\section{Lagrangian}
\indent

 Let the electromagnetic field couple to charges $q_0$ and $q_\pi$ at the ends of the open string. The interaction Lagrangian is given by
\begin{equation}
\label{b1}
 L_I=q_0\dot{X}^\mu A_\mu (\sigma = 0 )+q_\pi \dot{X}^\mu A_\mu (\sigma = \pi ).
\end{equation}
We choose a gauge, $A_\mu = -(1/2)B_{\mu \nu }X^{\nu }$, where $B_{\mu \nu }$ is a constant background field. By introducing a function $\rho (\sigma )$ such that
\begin{equation}
\label{b2}
 \rho (0) = q_0, \quad \rho (\pi ) = -q_\pi ,
\end{equation}
Eq.(2.1) can be written as
\begin{eqnarray}
\label{b3}
 L_I &=& \frac{1}{2} \rho (\sigma )\dot{X}BX(\sigma )\mid _{\sigma =0}^{\sigma =\pi }=
 \frac{1}{2}\int ^{\pi }_{0} d\sigma \partial _\sigma \{\rho (\sigma )\dot{X}BX(\sigma )\} \nonumber \\
        &=& \frac{1}{2}\int _0^{\pi}d\sigma (\rho '\dot{X}BX + \rho \dot{X}'BX + \rho \dot{X}BX') \\
        &=& \frac{1}{2}\int _0^{\pi}d\sigma \{ \rho '\dot{X}BX + 2\rho \dot{X}BX' + \partial _\tau (\rho X'BX) \}. \nonumber
\end{eqnarray}
Dropping the last total derivative term, we have the total Lagrangian
\begin{equation}
\label{b4}
 L = \frac{1}{4}\int _0^{\pi}d\sigma \{ \dot{X}^2 - X'^2 + 2\rho \dot{X}BX' +  \rho ' \dot{X}BX \}.
\end{equation}
where the coefficient is chosen as $1/4$ rather than the conventional 1/2 so that
the final results coincide with those of the conventional (\textit{i.e.} non-Dirac) quantization. (See Chap. IV and Appendix B.) Our system does not depend on the functional form of $\rho (\sigma )$ for $0< \sigma <\pi $. This can be seen from the fact that the action based on (\ref{b4}) is invariant under a variation with respect to $\rho (\sigma )$. 
The equation of motion and the boundary conditions follow from (\ref{b4})
\begin{eqnarray}
\label{b5}
 \ddot{X} - X'' = 0, \\
\label{b6}
 (X' + \rho B \dot{X})\mid _{0,\pi } = 0.
\end{eqnarray}
The canonical conjugate momentum is given by
\begin{equation}
\label{b7}
 P_c = \frac{1}{2}(\dot{X} + \rho BX' + \frac{1}{2} \rho ' BX).
\end{equation}

    Let us define an antisymmetric tensor $\beta ^\mu _{\ \nu}(\sigma )$ by
\begin{equation}
\label{b8}
 \rho (\sigma )B^\mu _{\ \nu }= \{ \tanh \beta (\sigma )\}^{\mu }_{\nu},
\end{equation}
and
\begin{equation}
\label{b9}
 \beta _0 = \beta (0), \quad \beta _\pi = \beta (\pi ),
\end{equation}
\begin{equation}
\label{b10}
 \rho(0)B = \tanh \beta _0, \qquad \rho (\pi )B = \tanh \beta _\pi .
\end{equation}
We shall choose $\beta (\sigma )$ to be linear in $\sigma $ as 
\begin{equation}
\label{b11}
 \beta (\sigma ) = \beta _0 - \gamma \sigma , \quad \gamma \equiv \frac{1}{\pi}(\beta _0 - \beta _\pi).
\end{equation}
The $\beta(\sigma)$ can be regarded as a "rotational angle" between $(\dot{X}, X')$ and 
$(P, Q)$, as is seen from Eqs.(7.5)-(7.8).  Hence, the Virasoro operator (\ref{h1}) does not depend on $\beta (\sigma $), as a result.  $\beta _0$, $\beta _\pi$, $\beta (\sigma )$   and $\gamma $ are all functions of $B$ and commutable with each other. In the following we extend the region of $\sigma $ to $(-\infty , \infty )$, but the physical region is still $(0,\pi )$.

\section{Constraints}
\setcounter{equation}{0}
\indent

 Define $\phi _l(\sigma )$, $\psi _l(\sigma )$ $(l = 0, \pi )$ by
 \begin{eqnarray}
\label{c1}
 \phi _l(\sigma ) \equiv \cosh \beta_l \cdot X'(\sigma ) + \sinh \beta_l \cdot \dot{X}(\sigma ), 
 \qquad (l = 0, \pi ) \\
\label{c2}
 \psi _l(\sigma ) \equiv \sinh \beta_l \cdot X'(\sigma ) + \cosh \beta_l \cdot \dot{X}(\sigma ). 
 \qquad (l = 0, \pi ) 
\end{eqnarray}
The boundary conditions (\ref{b6}) are expressed as
\begin{equation}
\label{c3}
 \phi _l(\sigma _l) \equiv \cosh \beta_l \cdot X'(\sigma _l) + \sinh \beta_l \cdot \dot{X}
 (\sigma _l) = 0. \qquad (\sigma_0=0, \sigma_\pi=\pi ) 
\end{equation}
These are primary constraints. From the consistency postulate
\begin{equation}
\label{c4}
 (\phi _l(\sigma _l), H)_{P.B} = 0, 
\end{equation}
we have the secondary constraints
\begin{eqnarray}
\label{c5}
 \cosh \beta_l \cdot \dot{X}'(\sigma _l) + \sinh \beta_l \cdot \ddot{X}(\sigma _l) 
 &=&\cosh \beta_l \cdot \dot{X}'(\sigma _l) + \sinh \beta_l \cdot X''(\sigma _l) \nonumber \\
 &=& \psi _l'(\sigma _l) = 0. 
\end{eqnarray}
 In the same way we get
\begin{equation}
\label{c6}
 \phi _l^{\ 2n}(\sigma _l) = \psi _l^{\ 2n+1}(\sigma _l) = 0. \quad (n=0,1,2,\cdots )
\end{equation}
Eqs.(\ref{c6}) show that $\phi _l(\sigma )$ ($\psi _l(\sigma )$) is the odd (even) function of 
$\sigma - \sigma_l$, that is,
\begin{equation}
\label{c7}
 \phi _l(2\sigma _l-\sigma) = -\phi _l(\sigma ), \quad \psi _l(2\sigma _l - \sigma) =
  \psi _l(\sigma).
\end{equation}
Let us write Eqs.(3.7) as
\begin{equation}
\label{c8}
 (1-\Delta _l)\psi _l = (1+\Delta _l)\phi _l = 0,
\end{equation}
where
\begin{equation}
\label{c9}
 (\Delta _lf)(\sigma ) = f(2\sigma _l - \sigma).
\end{equation}
Eqs.(3.8) are equivalent to one equation
\begin{eqnarray}
 (1-\Delta _l)\psi _l + (1+\Delta _l)\phi _l  &=& \psi _l + \phi _l - \Delta _l(
 \psi _l - \phi _l) \nonumber \\
\label{c10}
 &=& e^{\beta _l}(\dot{X} + X') - e^{-\beta _l}(\dot{X} - X') = 0. 
\end{eqnarray}
Useful formulas are
\begin{equation}
\label{c11}
 \Delta _l^{\ 2} = 1
\end{equation}
and
\begin{equation}
\label{c12}
 \textit{\textbf{I}}f(\sigma ) = \Delta _0f(2\pi - \sigma ) = f(2\pi +\sigma ),
\end{equation} 
where
\begin{equation}
\label{c13}
 \textit{\textbf{I}} \equiv \Delta _0\Delta _\pi, \qquad \textit{\textbf{I}}^{-1} = \Delta _\pi\Delta _0,
\end{equation}
I being the 2$\pi $-displacement operator.
\section{Poisson brackets}
\setcounter{equation}{0}
\indent 
 
The Dirac quantization will result in the commutator $[X(\sigma), P_c(\sigma')]$ to be $(1/2)$ 
$\delta (\sigma -\sigma ')$ for $0 < \sigma , \sigma ' < \pi$. This means that the correct momentum (the translation generator) is not $P_c$ but   
\begin{equation}
\label{d1}
 P \equiv  2P_c= \dot{X} + \rho BX' + \frac{1}{2}\rho 'BX .
\end{equation}
for which the Poisson bracket is given as
\begin{equation}
\label{d2}
(X(\sigma ), P(\sigma '))_{PB} = 2\delta (\sigma - \sigma ').
\end{equation}
This is the reason why we have chosen the factor $1/4$ in the Lagrangian (\ref{b4}). \\
\indent
 Let us define $P_\pm $ by
\begin{equation}
\label{d3}
 P_\pm = \frac{1}{2}(\dot{X} \pm X') = \frac{1}{2}(\hat{P} - \rho BX' \pm X')
\end{equation}
with
\begin{equation}
\label{d4}
 \hat{P} \equiv P - \frac{1}{2}\rho'BX.
\end{equation}
Then Eq.(\ref{c10}) can be written as
\begin{equation}
\label{d5}
 \chi \equiv P_+ - e^{-2\beta _l}\Delta _lP_- = 0.
\end{equation}
Poisson brackets for $X, X'$ and $\hat{P}$ are
\begin{equation}
\label{d6a}
 (X(\sigma), \hat{P}(\sigma'))_{PB} = 2\delta (\sigma - \sigma'), 
\end{equation}
\begin{equation}
\label{d6b}
 (X'(\sigma), \hat{P}(\sigma'))_{PB} = 2\delta '(\sigma - \sigma'), 
\end{equation}
\begin{equation}
\label{d6c} 
 (\hat{P}(\sigma), \hat{P}(\sigma'))_{PB} = -2\rho'(\sigma)B\delta (\sigma - \sigma'),
\end{equation}
where $\delta $ is the ordinary (non-periodic) delta function. We express (\ref{d6a}), (\ref{d6b}) and (\ref{d6c}) symbolically as
\begin{equation}
\label{d7a}
 (X, \hat{P})_{PB} = 2,
\end{equation}
\begin{equation}
\label{d7b}
 (X', \hat{P})_{PB} = 2\partial ,
\end{equation}
\begin{equation}
\label{d7c}
 (\hat{P}, \hat{P})_{PB} = -2\rho'B = -2\beta '\cdot \textrm{sech}^2\beta .
 \nonumber \\ 
\end{equation}
From these Eqs. we have
\begin{equation}
\label{d8}
 (P_\pm , P_\pm )_{PB} = \pm\partial, \quad (P_\pm , P_\mp )_{PB} = 0.
\end{equation}
Noting 
\begin{equation}
\label{d9}
 \partial\Delta _l = -\Delta _l\partial,
\end{equation}
we obtain
\begin{eqnarray}
 G_{ll'} &\equiv& (\chi _l,\chi _{l'})_{PB} = \partial - 
 e^{-2\beta _l+2\beta _{l'}}\Delta _l\partial\Delta _{l'} =
  (1+e^{-2\beta _l+2\beta _{l'}}\Delta _l\Delta _{l'})\partial \nonumber \\
\label{d10}
 &\equiv &H_{ll'}\partial , 
\end{eqnarray}
 where
\begin{equation}
\label{d11}
 H \equiv \left(
  \begin{array}{cc}
    2   &  1+e^{-2\pi\gamma}\textit{\textbf{I}}  \\
    1+e^{2\pi\gamma}\textit{\textbf{I}}^{-1}   &  2  \\
  \end{array}
\right). \quad \beta _0 - \beta _\pi=\pi\gamma
\end{equation}

      In order to define the Dirac bracket, we need the inverse of $G$ in its non-singular subspace. In the subspace $\textit{\textbf{I}} = e^{2\pi\gamma}$ the rank of $H$ is reduced, that is, $H$ becomes 
\begin{equation}
\label{d12}
 H = \left(
  \begin{array}{cc}
    2   &  2  \\
    2   &  2  \\
  \end{array}
\right) = 2\left(
  \begin{array}{c}
    1   \\
    1   \\
  \end{array}
\right) \left(
  \begin{array}{cc}
   1,    &  1  \\
  \end{array}
\right) = 4uu^T, \quad u=\frac{1}{\sqrt{2}}\left(
  \begin{array}{c}
    1   \\
    1   \\
  \end{array}
\right).
\end{equation}
Hence, defining the projection operator to the subspace $I = e^{2\pi\gamma}$ by $\Lambda_{\gamma}$, 
and $\bar{\Lambda }_{\gamma} \equiv 1- \Lambda_{\gamma}$ , we have
\begin{equation}
\label{d13}
 G^{-1} = \partial ^{-1}(\bar{\Lambda }_{\gamma}H^{-1} + \frac{1}{4}\Lambda _{\gamma}uu^T) 
 = (\bar{\Lambda }_{\gamma}H^{-1} + \frac{1}{4}\Lambda _{\gamma}uu^T)\partial ^{-1}. 
\end{equation}
The first term is written as
\begin{equation}
\label{d14}
 \bar{\Lambda }_{\gamma}H^{-1} = \bar{\Lambda }_{\gamma}D_{\gamma}^{\ -1}\left(
  \begin{array}{cc}
    2   &  -1-e^{-2\pi\gamma}\textit{\textbf{I}}  \\
    -1-e^{2\pi\gamma}\textit{\textbf{I}}^{-1}   &  2  \\
  \end{array}
\right),
\end{equation}
where
\begin{equation}
\label{d15}
 D_{\gamma} = 4 - (1+e^{-2\pi\gamma}\textit{\textbf{I}})(1+e^{2\pi\gamma}\textit{\textbf{I}}^{-1}) 
 = (1-e^{-2\pi\gamma}\textit{\textbf{I}})(1-e^{2\pi\gamma}\textit{\textbf{I}}^{-1}).
\end{equation}
There is a constant term in $\partial ^{-1}$
\begin{equation}
\label{d16}
 \langle \sigma| \partial  ^{-1}|\sigma'\rangle = \frac{1}{2}\epsilon (\sigma - \sigma ') + c, 
\end{equation}
where $\epsilon (\sigma)=$sgn$(\sigma)$.

     The Dirac bracket is then given by
\begin{equation}
\label{d17}
 (A,B)_{DB} = (A,B)_{PB} - \sum _{l=0,\pi }\sum _{l'=0,\pi }(A,\chi _l)_{PB}(G^{-1})_{ll'}
 (\chi _{l'},B)_{PB}.
\end{equation}
 $\Delta _l$ and $\textit{\textbf{I}}$ satisfy the following relations:
\begin{equation}
\label{d18}
 \Delta _0\textit{\textbf{I}} = \Delta _\pi, \qquad \textit{\textbf{I}}\Delta _\pi = \Delta _0, \qquad 
 \Delta _\pi\textit{\textbf{I}}^{-1} = \Delta _0, \qquad \textit{\textbf{I}}^{-1}\Delta _0 = \Delta _\pi, 
\end{equation}
\begin{equation}
\label{d19}
 \Delta _l\textit{\textbf{I}} = \textit{\textbf{I}}^{-1}\Delta _l.
\end{equation}
From (\ref{d19}) one can see for the projction $\Lambda _\gamma$
\begin{equation}
\label{d20}
 \Delta _l\Lambda _\gamma= \Lambda _{-\gamma}\Delta _l.
\end{equation}
 
\section{Dirac brackets}
\setcounter{equation}{0}
\indent

Here we use $\beta (\sigma )$ defined by (\ref{b11}). The operation $\Delta _l$ to $\beta (\sigma)$  gives
\begin{equation}
\label{e1}
 \Delta _l\beta (\sigma) = \beta (2\sigma_l-\sigma)=2\beta_l-\beta(\sigma).
\end{equation}
Since from (4.6) we have
\begin{equation}
\label{e2}
 (X, P_{\pm })_{PB}=1,
\end{equation}
\begin{equation}
\label{e3}
 (\hat{P}, P_{\pm })_{PB}=-\rho'B+\partial \rho B \pm\partial = \rho B\partial \pm\partial 
 =\pm \textrm{sech}\beta\cdot e^{\pm\beta}\partial,
\end{equation}
it follows that
\begin{equation}
\label{e4}
 (X, \chi _l)_{PB}=1-e^{2\beta _l}\Delta _l, 
\end{equation}
\begin{equation}
\label{e5}
 (\cosh\beta\cdot \hat{P}, \chi _l)_{PB}=e^\beta\partial+e^{-\beta}\partial e^{2\beta_l}\Delta_l
 =(e^\beta-e^{-\beta}e^{2\beta_l}\Delta_l)\partial.
\end{equation}
Then, defining $\tilde{P}$ by
\begin{equation}
\label{e6}
 \tilde{P} \equiv \cosh\beta \cdot \hat{P}, 
\end{equation}
we get, from (\ref{e1}), a formula
\begin{equation}
\label{e7}
 (\tilde{P}, \chi _l)_{PB} = (1 - \Delta_l)e^\beta\partial.
\end{equation}
In the following, by using Eqs.(\ref{e4}), (\ref{e7}) and (\ref{d17}) we calculate Dirac brackets for $X$ and $\tilde{P}$. \\
\\
\noindent
1. $(X, X)_{DB}$ \\
\indent
 First let us calculate
\begin{eqnarray}
 (X,X)_{DB} &=& -\sum _{l,l'}(X,X_l)_{PB}(G^{-1})_{ll'}(\chi _{l'},X)_{PB} \nonumber \\
 \label{e8}
 &=& \sum _{l,l'}(1-e^{2\beta_l}\Delta_{l})\partial ^{-1}
 (\bar{\Lambda}_\gamma H^{-1} 
 +\frac{1}{4}\Lambda _\gamma uu^T)_{ll'}(1-e^{-2\beta_{l'}}\Delta_{l'}).  
\end{eqnarray}
The contributions from the constant term of $\partial ^{-1}$ can be shown to be $2c$, if one notices that $\Delta _l=\textit{\textbf{I}}=1$ for any constant function. The other term $(1/2)\epsilon $ anti-commutes with $\Delta _l$ since $\epsilon $ is an odd function. The $\bar{\Lambda }_\gamma$ term becomes, by using Eqs.(\ref{d14}), (\ref{d15}) and (\ref{d18}), as follows:
\begin{equation}
\label{e9}
 \sum _{l'}\bar{\Lambda }_\gamma(H^{-1})_{ll'}(1-e^{-2\beta _{l'}}\Delta _{l'})
 = \bar{\Lambda }_\gamma(1-e^{-2\pi\gamma}\textit{\textbf{I}})^{-1}\left(
  \begin{array}{c}
    -e^{-2\pi\gamma}\textit{\textbf{I}}- e^{-2\beta _0}\Delta _0  \\
    1 + e^{-2\beta _0}\Delta _0   \\
  \end{array}
\right).
\end{equation}
Also, from (\ref{d20})
\begin{equation}
\label{e10}
 (1-e^{-2\beta _{l}}\Delta _{l})\bar{\Lambda }_\gamma
 =\bar{\Lambda }_\gamma- \bar{\Lambda }_{-\gamma}e^{-2\beta _l}\Delta _l,
\end{equation}
and the contribution from $(1/2)\epsilon $ is
\begin{eqnarray}
\label{e11}
 \left(
  \begin{array}{cc}
   \bar{\Lambda }_\gamma-\bar{\Lambda }_{-\gamma}e^{2\beta _0}\Delta _0 , &
   \bar{\Lambda }_\gamma-\bar{\Lambda }_{-\gamma}e^{2\beta _\pi}\Delta _\pi    \\
  \end{array}
\right) (1-e^{-2\pi\gamma}\textit{\textbf{I}})^{-1}\left(
  \begin{array}{c}
    -e^{-2\pi\gamma}\textit{\textbf{I}} + e^{-2\beta _0}\Delta _0   \\
    1 - e^{-2\beta _0}\Delta _0   \\
  \end{array}
\right) \frac{\epsilon }{2} \nonumber \\
 = \bar{\Lambda }_{\gamma}(1-e^{-2\pi\gamma}\textit{\textbf{I}})^{-1}
 (1-e^{-2\pi\gamma}\textit{\textbf{I}})\frac{\epsilon }{2} -
 \bar{\Lambda }_{-\gamma}(1-e^{-2\pi\gamma}\textit{\textbf{I}}^{-1})^{-1}
 (1-e^{-2\pi\gamma}\textit{\textbf{I}}^{-1})\frac{\epsilon }{2} \nonumber \\
 = (\bar{\Lambda }_{\gamma} - \bar{\Lambda }_{-\gamma})\frac{\epsilon }{2}
 = -(\Lambda _{\gamma} - \Lambda _{-\gamma})\frac{\epsilon }{2}.
\end{eqnarray}
 Next, the $\Lambda _{\gamma}$ terms are
\begin{equation}
\label{e12a}
 \Lambda _{\gamma}e^{-2\beta _\pi }\Delta _\pi 
 = \Lambda _{\gamma}e^{2\pi\gamma}\textit{\textbf{I}}^{-1}e^{-2\beta _0}\Delta _0
 = \Lambda _{\gamma}e^{-2\beta _0}\Delta _0,
\end{equation}
\begin{equation}
\label{e12b}
 e^{2\beta _\pi }\Delta _\pi \Lambda _{\gamma} 
 = e^{2\beta _0}\Delta _0e^{-2\pi\gamma}\textit{\textbf{I}}\Lambda _{\gamma} 
 = e^{2\beta _0}\Delta _0\Lambda _{\gamma}.
\end{equation}
Hence
\begin{eqnarray}
\label{e13}
 \frac{1}{2}(1-e^{2\beta _0}\Delta _0)\Lambda _{\gamma}(1+e^{-2\beta _0}
 \Delta _0)\frac{\epsilon }{2} 
 = \frac{1}{2}(\Lambda _{\gamma}-\Lambda _{-\gamma}e^{2\beta _0}\Delta _0)
 (1+e^{-2\beta _0}\Delta _0)\frac{\epsilon }{2} \nonumber \\
 = \frac{1}{2}\{\Lambda _{\gamma}(1+e^{-2\beta _0}\Delta _0) -
 \Lambda _{-\gamma}(1+e^{2\beta _0}\Delta _0)\}\frac{\epsilon }{2}.
\end{eqnarray}
Collecting all together we finally obtain
\begin{equation}
\label{e14}
 (X,X)_{DB} = -\frac{1}{2}\{\Lambda _{\gamma}(1-e^{-2\beta _0}\Delta _0) -
 \Lambda _{-\gamma}(1-e^{2\beta _0}\Delta _0)\}\frac{1}{2}\epsilon  
 + 2c|s\rangle \langle s|,
\end{equation}
where
\[
 |s\rangle \equiv \int _{-\infty }^{\infty } d\sigma |\sigma \rangle , 
 \qquad \langle s|s\rangle = 1.
\]
\\
\noindent
2. $(\tilde{P}, X)_{DB}$ \\
\indent
 This is given by
\begin{eqnarray}
\label{e15}
 (\tilde{P},X)_{DB} &=& (\tilde{P},X)_{PB} - \sum _{ll'}(\tilde{P},\chi _l)_{PB}
 (G^{-1})_{ll'}(\chi _{l'},X)_{PB} \nonumber \\
 &=& -2\cosh\beta + \sum _{ll'}(1-\Delta _l)e^\beta 
 (\bar{\Lambda }_{\gamma}H^{-1} + \frac{1}{4}\Lambda _{\gamma}uu^T)_{ll'}
 (1-e^{-2\beta _{l'}}\Delta _{l'}).
\end{eqnarray}
The use of (\ref{e1}) results in
\begin{equation}
\label{e16}
 \textit{\textbf{I}}\beta = (\beta - 2\pi\gamma)\textit{\textbf{I}},
\end{equation}
hence
\begin{eqnarray}
\label{e17}
 e^\beta e^{-2\pi\gamma}\textit{\textbf{I}} = \textit{\textbf{I}}e^\beta , \nonumber \\
\label{e18}
 e^\beta \Lambda _\gamma= \Lambda _0e^\beta
\end{eqnarray}
and
\begin{equation}
\label{e19}
 \Delta _l(1-\textit{\textbf{I}})^{-1}=(1-\textit{\textbf{I}}^{-1})^{-1}\Delta _l
 =-(1-\textit{\textbf{I}})^{-1}\textit{\textbf{I}}\Delta _l.
\end{equation}
These relations serve to calculate the $\bar{\Lambda }_\gamma$ term in (\ref{e15}) 
\begin{equation}
\label{e20}
 \bar{\Lambda }_0(1-\textit{\textbf{I}})^{-1}\left(
  \begin{array}{cc}
   1+\Delta _0\textit{\textbf{I}}^{-1},  &  1+\Delta _0\  \\
  \end{array}
\right)\left(
  \begin{array}{c}
    -\textit{\textbf{I}}e^\beta - \Delta _0e^{-\beta }   \\
    e^\beta + \Delta _0e^{-\beta }   \\
  \end{array}
\right) = 2\bar{\Lambda }_0\cosh \beta .
\end{equation}
and the $\Lambda _\gamma$ term
\begin{equation}
\label{e21}
 \frac{1}{2}\Lambda _0(1-\Delta _0)(e^\beta - \Delta _0e^{-\beta })
 = \Lambda _0(1-\Delta _0)\cosh \beta .
\end{equation}
Thus we get 
\begin{equation}
\label{e22}
 (\tilde{P},X)_{DB} = -\Lambda _0(1+\Delta _0)\cosh \beta .
\end{equation}
\\
\noindent
3. $(\tilde{P}, \tilde{P})_{DB}$ \\
\indent
 This is given by
\begin{eqnarray}
\label{e23}
 (\tilde{P}, \tilde{P})_{DB} &=& -2\beta ' -\sum _{ll'}(\tilde{P},\chi _l)_{PB}
 (G^{-1})_{ll'}(\chi _{l'},\tilde{P})_{PB} \nonumber \\
 &=& -2\beta ' -\sum _{ll'}(1-\Delta _l)e^\beta 
 (\bar{\Lambda }_\gamma H^{-1}+\frac{1}{4}\Lambda _\gamma uu^T)_{ll'}
 \partial e^{-\beta }(1-\Delta _{l'}) \nonumber \\
 &=& -2\beta ' -\sum _{ll'}(1-\Delta _l)
 (\bar{\Lambda }_0e^\beta H^{-1}+\frac{1}{4}\Lambda _0e^\beta uu^T)_{ll'}
 \partial e^{-\beta }(1-\Delta _{l'}).
\end{eqnarray}
The $\bar{\Lambda }_\gamma$ term becomes
\begin{eqnarray}
\label{e24}
 -\bar{\Lambda }_0(1-\Delta_0,1-\Delta_\pi)e^\beta D_\gamma ^{\ -1}\left(
  \begin{array}{cc}
    2   &  -1-e^{-2\pi\gamma}\textit{\textbf{I}}  \\
    -1-e^{2\pi\gamma}\textit{\textbf{I}}^{-1}   & 2   \\
  \end{array}
\right)\partial e^{-\beta }\left(
  \begin{array}{c}
    1-\Delta_0   \\
    1-\Delta_\pi   \\
  \end{array}
\right) \nonumber \\
  = -\bar{\Lambda }_0D_0^{\ -1}\left(
  \begin{array}{cc}
   1-\Delta_0,    &  1-\Delta_\pi  \\
  \end{array}
\right)\left(
  \begin{array}{cc}
    2   &  -1-\textit{\textbf{I}}  \\
    -1-\textit{\textbf{I}}^{-1}   &  2  \\
  \end{array}
\right)e^\beta \partial e^{-\beta }\left(
  \begin{array}{c}
    1-\Delta_0   \\
    1-\Delta_\pi   \\
  \end{array}
\right) 
\end{eqnarray}
and
\begin{equation}
\label{e25}
 e^\beta \partial e^{-\beta}\left(
  \begin{array}{c}
    1-\Delta_0   \\
    1-\Delta_\pi   \\
  \end{array}
\right) = (\partial -\beta ')\left(
  \begin{array}{c}
    1-\Delta_0   \\
    1-\Delta_\pi   \\
  \end{array}
\right) = \left(
  \begin{array}{c}
    1+\Delta_0   \\
    1+\Delta_\pi   \\
  \end{array}
\right)\partial - \left(
  \begin{array}{c}
    1-\Delta_0   \\
    1-\Delta_\pi   \\
  \end{array}
\right)\beta '.
\end{equation}
Hence, the $\bar{\Lambda }_\gamma$ term results in
\begin{equation}
\label{e26}
 2\bar{\Lambda }_0\beta '.
\end{equation}
The other $\Lambda _\gamma$ term becomes
\begin{eqnarray}
\label{e27}
 -\frac{1}{2}\Lambda _0(1-\Delta _0)e^\beta \partial e^{-\beta}(1-\Delta_0)
 &=& -\frac{1}{2}\Lambda _0(1-\Delta _0)(\partial -\beta')(1-\Delta_0) \nonumber \\
 &=& \Lambda _0(1-\Delta _0)\beta'.
\end{eqnarray}
Finally we have
\begin{equation}
\label{e28}
 (\tilde{P},\tilde{P})_{DB} = -\Lambda _0(1+\Delta _0)\beta'.
\end{equation}

\section{Quasi-periodic space}
\setcounter{equation}{0}
\indent

 In this section we give a $\sigma$-dependence of the Dirac brackets. \\
\indent
 Let the subspace $\textit{\textbf{I}}=e^{2\pi\gamma}$ be $V_\gamma $, which is projected by $\Lambda _\gamma$. This is a function space satisfying the quasi-periodicity 
\begin{equation}
\label{f1} 
 f(\sigma+2\pi) = e^{2\pi\gamma}f(\sigma).
\end{equation}
The element $f$ of this space can be expanded into the Fourier-like series
\begin{eqnarray}
\label{f2}
 f(\sigma) = \sum _{n}e^{(\gamma - in)\sigma}f_n, \quad (f \in V_\gamma) \\
 f_n = \frac{1}{2\pi }\int _{-\pi}^\pi d\sigma e^{-(\gamma - in)\sigma}f(\sigma). 
 \quad (f \in V_\gamma)
\end{eqnarray}
Therefore, we have the expression
\begin{eqnarray}
\label{f3}
 \langle \sigma |\Lambda _\gamma|\sigma '\rangle 
 &=&\frac{1}{2\pi }\sum _{n}e^{(\gamma - in)(\sigma-\sigma')}
 =e^{\gamma(\sigma-\sigma')}\bar{\delta}(\sigma-\sigma') \nonumber  \\
 &=&\sum _{k}e^{2k\pi\gamma}\delta (\sigma-\sigma'-2k\pi),
\end{eqnarray}
where $\bar{\delta }$ is the periodic delta function.
As for any function $f$ which does not belong to $V_\gamma$ , we get, using the Fourier integral
\begin{equation}
\label{f4} 
 f(\sigma)=\int _{-\infty}^{\infty}dk e^{-ik\sigma}\tilde{f}(k),
\end{equation}
the following expansion:
\begin{eqnarray}
\label{f5}
 \langle \sigma |\Lambda _\gamma|f\rangle 
 &=&\int _{-\infty}^{\infty}d\sigma'\langle \sigma |\Lambda _\gamma|\sigma'\rangle
  \int _{-\infty}^{\infty}dk e^{-ik\sigma'}\tilde{f}(k) \nonumber \\
 &=&\sum _{n}e^{(\gamma - in)\sigma}\int _{-\infty}^{\infty}dk\frac{1}{2\pi}
    \int _{-\infty}^{\infty}d\sigma'e^{-i(k-n-i\gamma)\sigma'}\tilde{f}(k) \nonumber \\
 &=&\sum _{n}e^{(\gamma - in)\sigma}\int _{-\infty}^{\infty}dk\delta (k-n-i\gamma)\tilde{f}(k)
\end{eqnarray}
that is,
\begin{equation}
\label{f6}
 \langle \sigma |\Lambda _\gamma|f\rangle 
 =\sum _{n}e^{(\gamma - in)\sigma}\tilde{f}(n+i\gamma).
\end{equation}
Here we have used a fact that eigenvalues of $i\gamma $ are real for space components
of $\gamma $. ( As for time components of $\gamma $, it may be necessary to consider Euclidization.)

\indent
 As for $\Lambda _\gamma\epsilon $ in (5.14), we use the Fourier integral of 
 $\epsilon (\sigma)$
\begin{equation}
\label{f7}
 \epsilon (\sigma-\sigma')=\frac{i}{\pi}\int _{-\infty}^{\infty}dk e^{-ik(\sigma-\sigma')}
 \frac{k}{k^2+\epsilon ^2}
\end{equation} 
to give
\begin{eqnarray}
\label{f8}
 \langle \sigma |\Lambda _\gamma\epsilon |\sigma'\rangle 
 &=&\frac{1}{2\pi^2i}\sum _{n}e^{(\gamma - in)(\sigma-\sigma')}
 \int _{-\infty}^{\infty}dk\frac{k}{k^2+\epsilon ^2}
 \int _{-\infty}^{\infty}d\sigma''e^{i(k+n+i\gamma)\sigma''} \nonumber \\
 &=&\frac{i}{\pi}\sum _{n}e^{(\gamma - in)(\sigma-\sigma')}
 \frac{n+i\gamma}{(n+i\gamma)^2+\epsilon ^2}.
\end{eqnarray}
If the eigenvalue of $i\gamma $ is not integer, this defines the function 
\begin{equation}
\label{f9} 
 \bar{\epsilon }(\sigma-\sigma':\gamma)\equiv 
 \langle \sigma |\Lambda _\gamma\epsilon |\sigma'\rangle 
 = \frac{1}{\pi}\sum _{n}\frac{1}{\gamma-in}e^{(\gamma - in)(\sigma-\sigma')}.
\end{equation}
In terms of this function Eq.(\ref{e14}) can be written as
\begin{equation}
\label{f10}
 \langle \sigma |(X,X)_{DB}|\sigma'\rangle =(X(\sigma),X(\sigma'))_{DB}
 =-\frac{1}{4}E(\sigma,\sigma':\gamma)+2c,
\end{equation}
where
\begin{equation}
\label{f11}
 E(\sigma,\sigma':\gamma)\equiv \bar{\epsilon}(\sigma-\sigma':\gamma)
 +\bar{\epsilon }(-\sigma+\sigma':\gamma)
 +e^{-2\beta_0}\bar{\epsilon }(\sigma+\sigma':\gamma)
 +e^{2\beta_0}\bar{\epsilon }(-\sigma-\sigma':\gamma).
\end{equation}

\indent
 In (\ref{e22}) and (\ref{e28}) we have
\begin{equation}
\label{f12}
 \langle \sigma |\Lambda _0(1+\Delta _0)|\sigma'\rangle 
 =\bar{\delta }(\sigma-\sigma')+\bar{\delta}(\sigma+\sigma')=\delta_c(\sigma,\sigma'),
\end{equation}
hence
\begin{eqnarray}
\label{f13}
 (\tilde{P}(\sigma),X(\sigma'))_{DB}=-\delta_c(\sigma,\sigma')\cosh\beta(\sigma'), \\
\label{f14}
 (\tilde{P}(\sigma),\tilde{P}(\sigma'))_{DB}=-\delta_c(\sigma,\sigma')\beta'(\sigma')
 =\gamma\delta_c(\sigma,\sigma').
\end{eqnarray}
\indent
 The quantized theory is given by replacing the Dirac bracket with the commutator
\begin{equation}
\label{f15}
 (A,B)_{DB} \  \to \ -i[A,B].
\end{equation}
We summarize relevant commutation relations:
\begin{equation}
\label{f17}
 [X(\sigma),\tilde{P}(\sigma')]=i\cosh\beta(\sigma)\delta _c(\sigma,\sigma'), 
\end{equation}
\begin{equation}
\label{f18}
 [\tilde{P}(\sigma),\tilde{P}(\sigma')]=i\gamma\delta _c(\sigma,\sigma'), 
\end{equation}
\begin{equation}
\label{f20}
 [X(\sigma),P(\sigma')]=i\{\delta _c(\sigma,\sigma')-
 \frac{1}{2}\gamma \textrm{sech}^2\beta(\sigma)[X(\sigma),X(\sigma')]\}, 
\end{equation}
\begin{equation}
\label{f21}
 [P(\sigma),P(\sigma')]=\gamma \textrm{sech}^2\beta(\sigma)[X(\sigma),X(\sigma')]
 \gamma \textrm{sech}^2\beta(\sigma'), 
\end{equation}
\begin{equation}
\label{f22}
 [X(\sigma),X(\sigma')]=-\frac{i}{4}E(\sigma,\sigma':\gamma)+i2c. 
\end{equation}
where
\begin{equation}
\label{f23}
 2c=\frac{\cosh\beta_0\cosh\beta_\pi}{\sinh\pi\gamma}.
\end{equation}
In the Appendix A we derive the explicit form of $E$ and give the proof of Eq.(\ref{f23}).  From Eq.(\ref{f22}) one finds noncommutativity of $X$ at both ends
\begin{eqnarray}
\label{f24}
 [X(0),X(0)]=-i\cosh\beta_0\sinh\beta_0, \\
\label{f25}
 [X(\pi),X(\pi)]=i\cosh\beta_\pi\sinh\beta_\pi.
\end{eqnarray}
According to these equations we see that noncommutativity reveals also to $[P, P]$ and $[X, P]$.

\section{Mode expansions}
\setcounter{equation}{0}
\indent

 In this section we consider mode expansions of relevant fields. \\
\indent
 The constraint (4.5) leads to
\begin{eqnarray}
\label{g1a}
 P_+ &=& e^{-2\pi\gamma}\textit{\textbf{I}}P_+ \qquad \in  V_\gamma, \\
\label{g1b}
 P_- &=& e^{2\beta_0}\Delta _0P_+ \qquad \in V_{-\gamma}.
\end{eqnarray}
Hence Fourier-like expansions of them are
\begin{eqnarray}
\label{g2a}
 P_+(\sigma) = \frac{1}{2\sqrt{\pi}}\sum _{n}e^{(\gamma-in)\sigma-\beta_0}\alpha _n, \\
\label{g2b}
 P_-(\sigma) = \frac{1}{2\sqrt{\pi}}\sum _{n}e^{-\{(\gamma-in)\sigma-\beta_0\}}\alpha _n.
\end{eqnarray}
From the definition (\ref{d3}) of $P_\pm$, therefore, we have mode expansions
\begin{eqnarray}
\label{g3a}
 \dot{X}(\sigma)=P_+(\sigma)+P_-(\sigma)
 &=&\frac{1}{\sqrt{\pi}}\sum _{n}\cosh\{(\gamma-in)\sigma-\beta_0\}\alpha _n \nonumber \\
 &=&\cosh\beta(\sigma)\cdot \tilde{P}(\sigma)-\sinh\beta(\sigma)\cdot \tilde{Q}(\sigma), \\
\label{g3b}
 X'(\sigma)=P_+(\sigma)-P_-(\sigma)
 &=&\frac{1}{\sqrt{\pi}}\sum _{n}\sinh\{(\gamma-in)\sigma-\beta_0\}\alpha _n \nonumber \\
 &=&-\sinh\beta(\sigma)\cdot \tilde{P}(\sigma)+\cosh\beta(\sigma)\cdot \tilde{Q}(\sigma),
\end{eqnarray}
where
\begin{eqnarray}
\label{g4a}
 \tilde{P}(\sigma)&=&\cosh\beta(\sigma)\cdot\dot{X}(\sigma)+\sinh\beta(\sigma)\cdot X'(\sigma)
 =\frac{1}{\sqrt{\pi}}\sum _{n}\cos n\sigma\cdot\alpha_n, \\
\label{g4b}
 \tilde{Q}(\sigma)&=&\sinh\beta(\sigma)\cdot\dot{X}(\sigma)+\cosh\beta(\sigma)\cdot X'(\sigma)
 =-i\frac{1}{\sqrt{\pi}}\sum _{n}\sin n\sigma\cdot\alpha_n. 
\end{eqnarray}
From the equation of motion (2.5) for $X$, one can see the $\tau $-dependence of $\alpha _n$
to be
\begin{equation}
\label{g5}
 \alpha _n(\tau )=e^{(\gamma-in)\tau }\alpha_n(0)=\alpha_n(0)e^{(-\gamma-in)\tau }.
\end{equation} 
The mode expansion of $X$, is, therefore, given by 
\begin{equation}
\label{g6}
 X(\tau,\sigma)=\frac{1}{\sqrt{\pi}}\sum _{n}\frac{1}{\gamma-in}e^{(\gamma-in)\tau}
 \cosh\{(\gamma-in)\sigma-\beta_0\}\alpha_n(0)+\frac{1}{\sqrt{\pi}}b,
\end{equation}
and
\begin{equation}
 P(\tau ,\sigma ) = \textrm{sech}\beta(\sigma)\tilde{P}(\tau ,\sigma)
 +\frac{1}{2}\gamma \textrm{sech}^2\beta (\sigma)X(\tau ,\sigma).
\end{equation}
The over-all time-development factor $e^{\gamma \tau }$ in (7.10) means that $i\gamma $ (or its positive eigenvalue) is the cyclotron frequency of the string.
The commutation relation for the mode operators $\alpha_n$ is found to be 
\begin{equation}
\label{g8}
 [\alpha_m, \alpha_n]=(m+i\gamma)\delta _{m+n,0},
\end{equation}
owing to (\ref{f17}) and (\ref{f18}).

\section{Virasoro algebra}
\setcounter{equation}{0}
\indent

 The Virasoro operator is defined by (see Appendix B for note)
\begin{equation}
\label{h1}
  L_n  = \frac{1}{4}\int _{-\pi}^{\pi}d\sigma e^{\pm in\sigma}:(\dot{X} \pm X')^2: 
  = \frac{1}{2}\sum _{k}:\alpha _k\cdot \alpha_{n-k}:,
\end{equation}
which satisfies the Virasoro algebra
\begin{equation}
\label{h3}
 [L_m,L_n] = (m-n)L_{m+n} + m(am^2-b)\delta _{m+n,0},
\end{equation}
where
\begin{equation}
\label{h4}
 a=\frac{1}{12}d, \qquad b=\frac{1}{12}d-\frac{1}{2}\textrm{tr}(\gamma ^2). \qquad (d=\textrm{space-time dimension})
\end{equation}
The central term differs from the conventional one owing to the unconventional commutation relation (\ref{g8}). This conforms the well-known result in Ref.[9].  The extra term  $-\textrm{tr}(\gamma^2)/2$ can be eliminated by a shift $L_0 \to L_0 + \textrm{tr}(\gamma^2)/4$.

\section{Dependence on the cyclotron frequency}
\setcounter{equation}{0}
\noindent
1. The periodicity on the parameters \\ 
\indent
In view of (2. 10) one sees that $\beta _0$ ($\beta _\pi$) may differ by an integral multiple of $i\pi$ to give the same $\rho (0)$ ($\rho (\pi)$).  This means that the system is $\pi$-periodic with respect to $i\beta_0$ and $i\pi\gamma$. In this section we would like to investigate how these periodicities are related to the mode operators. \\
\indent
First of all, Eq. (7. 10) tells us that the displacement of $\beta_0$ 
\begin{equation}
\label{i0a}
 \beta_0 \to \beta_0+in\pi 
\end{equation}
cause the change in the signature of $\alpha_n$:
\begin{equation}
\label{i0b}
 \alpha_n \to (-1)^n\alpha_n
\end{equation}
\indent
To investigate the periodicity in $\gamma$ it is convenient to represent $B$ in two-dimensional blocks.  For definiteness we consider the spatial components (related to the magnetic field). Then the $i$-th block has a form
\begin{equation}
\label{i1}
 B^{(i)}=\lambda ^{(i)}\epsilon =\lambda ^{(i)} \left(
  \begin{array}{cc}
    0   &  1  \\
    -1   &  0  \\
  \end{array}
\right).
\end{equation}
We represent also $\beta_0$, $\beta_\pi$ and $\gamma$  as
\begin{equation}
\label{i1b}
\beta_0^{(i)}=\pi\omega _0^{(i)}\epsilon , \qquad \beta_\pi^{(i)}=\pi\omega _\pi^{(i)}\epsilon , \qquad \gamma^{(i)}=\omega ^{(i)}\epsilon
\end{equation}
where
\begin{equation}
\label{i1c}
\omega^{(i)}= \omega _0^{(i)}-\omega _\pi^{(i)}
\end{equation}
is the cyclotron frequency mentioned in sect.VII.  Henceforth we confine ourselves to one particular block and suppress the index $i$. \\
\indent
 We adopt a complex basis in the two-dimensional subspace
\begin{equation}
\label{i2}
 e^\pm = \frac{1}{\sqrt{2}}(1,\pm i) = \frac{1}{\sqrt{2}}(e_1\pm ie_2),
\end{equation}
which diagonalizes $\epsilon $
\begin{equation}
\label{i3}
 \epsilon  \cdot e^\pm = \pm i e^\pm, \qquad e^\pm \cdot \epsilon  = \mp ie^\pm.
\end{equation}
Then from (7. 10) we have
\begin{eqnarray}
\label{i5}
 X^\pm(\tau,\sigma;\omega )&=&e^\pm\cdot X(\tau,\sigma;\omega) \nonumber \\
 &=&\frac{1}{\sqrt{\pi}}\sum _{n}\frac{i}{n\pm\omega }e^{-i(n\pm\omega )\tau}
 \cos\{(n\pm\omega )\sigma-\pi\omega _0\}\alpha_n^\pm(\omega )+\frac{1}{\sqrt{\pi}}b^\pm(\omega ).
\end{eqnarray}
We have explicitly written the $\omega $-dependence of the operators.  This expansion essentially coincides with that in Ref.[9]. \\
\indent
The mode operators in (9. 8) satisfy
\begin{equation}
\label{i7}
 \alpha _n^\pm(\omega)^\dagger = \alpha _{-n}^\mp(\omega), \qquad 
 b^\pm(\omega)^\dagger = b^\mp(\omega)
\end{equation}
and their non-vanishing commutators are
\begin{eqnarray}
\label{i8a}
 [\alpha_m^\pm(\omega ),\alpha_n^\mp(\omega )]&=&(m\pm\omega )\delta_{m+n,0}, \\
\label{i8b}
 [b^\pm(\omega ),b^\mp(\omega )]&=&-\pi\frac{\cos\pi\omega _0\cos\pi\omega _\pi}{\sin\pi\omega }.
\end{eqnarray}
\indent
 The periodicity on $\gamma$ requires $X^\pm(\tau,\sigma;\omega +l)=X^\pm(\tau,\sigma;\omega )$ for any integer $l$, which implies
 \begin{equation}
\label{i9}
 \alpha_n^\pm(\omega+l)=\alpha_{n\pm l}^\pm(\omega)
\end{equation}
and
\begin{equation}
\label{i10}
 b^\pm(\omega+l)=b^\pm(\omega).
\end{equation}
These relations are easily seen to be consistent with (9. 10) and (9. 11), if one notes the equality  $\omega_\pi=\omega_0-\omega$. The relation (9. 12) determines the $\omega$-dependence of the $\alpha_n^{\pm}(\omega)$'s as
\begin{equation}
\label{i11}
 \alpha_n^\pm(\omega)=\alpha_0^\pm(\omega\pm n).
\end{equation}
\noindent
2. The limit to the neutral string \\
\indent
 If $\omega $ tends to an integral value $m$, Eqs. (2. 11) and (2. 10) give $\rho(\pi)\to \rho(0)$ and we have the neutral string.  Let us see how this occurs for the mode operators. \\
\indent
 In Eq. (9. 8) the term with $n=\pm m$ becomes singular as $\omega \to m$.  This singularity should be cancelled by a similar singularity of $b^\pm(\omega)$. \\
\indent
 It is sufficient to consider the case $m=0$.  To retain the symmetry between the two end points we define
\begin{equation}
\label{i12}
 \bar{\beta}\equiv \frac{1}{2}(\beta_0+\beta_\pi), \qquad \beta_0=\bar{\beta}
 +\frac{1}{2}\pi\gamma, \qquad \beta_\pi=\bar{\beta}-\frac{1}{2}\pi\gamma.
\end{equation}
The commutation relations for $\alpha_0$ and $b$ are
\begin{eqnarray}
\label{i13}
 [\alpha_0, \alpha_0] = i\gamma, \\
\label{i14}
 [\alpha_0, b] = 0
\end{eqnarray}
and
\begin{equation}
\label{i15}
 [b, b]
 =i\gamma^{-1}\cosh^2\bar{\beta}+O(\gamma).
\end{equation}
\indent
 If we expand $\alpha_0$ as
\begin{equation}
\label{i16}
 \alpha_0^{\ i} \approx u^i+\frac{1}{2}(\gamma \cdot v)^i \qquad (\gamma \approx 0),
\end{equation}
from (9. 16) we get, 
\begin{equation}
\label{i17}
 [u^i, u^j]=0, \qquad [u^i, v^j]=-i\delta^{ij}.
\end{equation}
The $n = 0$ term in (7.10) is reduced to, apart from the factor $\frac{1}{\sqrt{\pi}}$, to
\begin{eqnarray}
 \gamma^{-1}e^{\gamma\tau}\cosh(\gamma\sigma-\beta_0)\alpha_0
  =\cosh\beta_0\cdot (\gamma^{-1}u+\tau u+\frac{v}{2})-\sigma\sinh\beta_0\cdot u+O(\gamma)
  \nonumber \\
\label{i18}
  =\cosh\bar{\beta}\cdot (\gamma^{-1}u+\tau u+\frac{v}{2})
  +(\frac{\pi}{2}-\sigma)\sinh\bar{\beta}\cdot u+O(\gamma).
\end{eqnarray}
On the other hand for the neutral string we have
\begin{eqnarray}
\label{i19}
 X(\tau,\sigma:\gamma=0)= x  
 +\frac{1}{\sqrt{\pi}}\{\tau \cosh\bar{\beta}+(\frac{\pi}{2}-\sigma )
 \sinh\bar{\beta }\}\cosh\bar{\beta }\cdot p \nonumber \\
 + \frac{1}{\sqrt{\pi}}\sum _{n\neq 0}\frac{i}{n}e^{-in\tau}
 \cos(n\sigma-i\bar{\beta})\alpha_n(\gamma=0)
\end{eqnarray}
where $(x, p)$ is the canonical pair.  It can be shown that the $X(\tau ,\sigma )$ in (7. 10) tends to (9. 22) and the commutation relations (9. 17) and (9. 18) hold in the lowest order in $\gamma$, if 
\begin{equation}
\label{i20}
 u=\frac{1}{\sqrt{\pi}}\cosh\bar{\beta}\cdot p, \qquad v=\sqrt{\pi}\textrm{sech}\bar{\beta}\cdot x
\end{equation}
and
\begin{equation}
\label{i21}
  b = -\cosh\bar{\beta}\cdot(\gamma^{-1}\cdot u-\frac{v}{2})+O(\gamma).
\end{equation}

\section{The $g$-factors of charged strings}
\setcounter{equation}{0}
\indent

 In this section we calculate the $g$-factor of the charged string.  The magnetic moment $\mbox{\boldmath $\mu $}$ of the system is defined, in the non-relativistic limit, as,
\begin{equation}
\label{j1}
 \mbox{\boldmath $\mu $} =-\frac{\partial E}{\partial \textit{\textbf{B}}}\Bigr|_{B=0} 
 =\frac{1}{2E}\frac{\partial p^2}{\partial \textit{\textbf{B}}}\Bigr|_{B=0}
 \approx \frac{1}{2M}\frac{\partial p^2}{\partial \textit{\textbf{B}}}\Bigr|_{B=0},
\end{equation}
where $E$ is the energy, $p$ the four-momentum and $M$ the mass of the particle. \\
\indent
 In the case of our charged string the four-momentum squared is fixed by the Virasoro on-shell condition
\begin{equation}
\label{j2}
 \{L_0-\alpha(0)\}|\Psi \rangle =0
\end{equation}
where
\begin{equation}
\label{j3}
 \alpha(0)=1-\frac{1}{4}\textrm{tr}(\gamma ^2)
\end{equation}
and
\begin{eqnarray}
 L_0&=&\frac{1}{2}\int _0^\pi d\sigma (\dot{X}^2+X'^2) \nonumber \\
\label{j4}
 &=&\frac{1}{2}\int _0^\pi d\sigma \{(P-\rho BX'-\frac{1}{2}\rho 'BX)^2+X'^2\}.
\end{eqnarray}
\indent
 For a weak magnetic field $\textit{\textbf{B}}$  Eq.(10. 4) becomes
\begin{equation}
\label{j5}
 \{L_0|_{B=0}-1-\textit{\textbf{B}}\cdot \textit{\textbf{G}}
 +O(\textit{\textbf{B}}^2)\}|\Psi \rangle =0,
\end{equation}
where
\begin{equation}
\label{j6}
 L_0|_{B=0}=(\sum _{n>0}\alpha_{-n}\cdot \alpha_n + \frac{1}{2}\alpha_0^2)|_{\textit{\textbf{B}}=0}
 =R+\frac{1}{2\pi}p^2
\end{equation}
and
\begin{equation}
\label{j7}
 \textit{\textbf{G}}=
 -\frac{\partial L_0}{\partial \textit{\textbf{B}}}\Bigr|_{B=0}
 =\textit{\textbf{G}}_1+\textit{\textbf{G}}_2
\end{equation}
with
\begin{equation}
\label{j8}
 \textit{\textbf{G}}_1=-\frac{1}{2}\int _0^\pi d\sigma (\rho '\textit{\textbf{X}}\times 
 \textit{\textbf{P}})\Bigr|_{B=0}
\end{equation}
and
\begin{equation}
\label{j9}
 \textit{\textbf{G}}_2=-\int _0^\pi d\sigma (\rho \textit{\textbf{X}}'\times 
 \textit{\textbf{P}})\Bigr|_{B=0}.
\end{equation}
\indent
 We consider the case that $|\Psi \rangle $ is a particle state with definite values of $R$ and $p^2$ .  The quantity $R$ is related to the mass operator $M$ of the particle through
\begin{equation}
\label{j10}
 R-1=\frac{1}{2\pi}M^2.
\end{equation}
We have therefore
\begin{equation}
\label{j11}
 [\frac{1}{2\pi}(p^2+M^2)-\textit{\textbf{B}}\cdot\textit{\textbf{G}}]|\Psi \rangle =0,
\end{equation}
which in view of Eq. (10. 1), gives
\begin{equation}
\label{j12}
 \mbox{\boldmath $\mu $}=\frac{\pi}{M}\textit{\textbf{G}}.
\end{equation}
\indent
 Before calculating $G$ we note that in the limit $\textit{\textbf{B}} \to 0$ the function $\rho (\sigma )$ becomes
\begin{equation}
\label{j13}
 \rho (\sigma )\approx \pi^{-1}\{\rho (\pi )-\rho(0)\}\sigma + \rho (0)
 =q_0-\pi^{-1}q\sigma
\end{equation}
with the total charge
\begin{equation}
\label{j14}
 q=q_0+q_\pi.
\end{equation}
\indent
 The quantity $\textit{\textbf{G}}_1$ is readily seen to be proportional to the total angular momentum
\begin{equation}
\label{j15}
 \textit{\textbf{G}}_1=\frac{1}{2\pi}q\textit{\textbf{J}}
 =\frac{1}{2\pi}q(\textit{\textbf{L}}+\textit{\textbf{S}})
\end{equation}
with the orbital and spin angular momenta
\begin{equation}
\label{j16}
 \textit{\textbf{L}}=\textit{\textbf{x}}\times \textit{\textbf{p}},
\end{equation} 
\begin{equation}
\label{j17}
 \textit{\textbf{S}}=
 \frac{i}{2}\sum _{n\neq 0}\frac{1}{n}\mbox{\boldmath $\alpha_n\times \alpha_{-n}$}.
\end{equation}
\indent
 The integration in (10. 9) can be performed to give
\begin{equation}
\label{j18}
 \textit{\textbf{G}}_2=
 \frac{i}{2\pi}\sum _{n\neq n'}\mbox{\boldmath$\alpha_n\times (\alpha_{n'}-\alpha_{-n'})$}\frac{q_0+(-)^{n-n'}q_\pi}{n-n'}.
\end{equation}
If the state $|\Psi \rangle $ is an eigenstate of $R$, only the terms of the form $\mbox{\boldmath$\alpha_n\times \alpha_{-n}$}$ contribute to the expectation value of $\textit{\textbf{G}}_2$:
\begin{equation}
\label{j19}
 \langle \Psi |\textit{\textbf{G}}_2|\Psi \rangle 
 =\frac{i}{2\pi}\sum _{n\neq 0}\frac{q_0+q_\pi}{2n}
 \langle \Psi |\mbox{\boldmath $\alpha_n\times \alpha_{-n}$}|\Psi \rangle
 =\frac{q}{2\pi}\langle \Psi |\textit{\textbf{S}}|\Psi \rangle . 
\end{equation}
\indent
 As a result we get
\begin{equation}
\label{j20}
 \langle \Psi |\mbox{\boldmath $\mu $}|\Psi \rangle 
 =\frac{q}{2M}\langle \Psi |(\textit{\textbf{L}}+2\textit{\textbf{S}})|\Psi \rangle
\end{equation}
so that the $g$-factors are the same as for the charged Dirac particle.
\begin{equation}
\label{j21}
 g_L=1 \qquad \textrm{and} \qquad g_S=2.
\end{equation}
\indent
 We note that any change of the function $\rho (\sigma ) \to \rho(\sigma)+\delta \rho (\sigma ) $ does not affect this result, since the expectation value of
\begin{equation}
\label{j22}
\delta\textit{\textbf{G}}=-\frac{1}{2}\int _0^\pi d\sigma \delta\rho (\textit{\textbf{X}}'\times  \textit{\textbf{P}} -
\textit{\textbf{X}}\times  \textit{\textbf{P}}') 
\end{equation}
is seen to vanish.

\section{Conclusions}
\setcounter{equation}{0}
\indent

 We have performed the Dirac quantization for constrained system of the charged string in the constant $B$ field. Noncommutativity appears at both ends of $X$ and also of $P$. We have considered the dependence of the cyclotron frequency of the charged string.  We have found that a displacement of the cyclotron frequency $\omega \to \omega +l $ ($l$: integer) causes the mode translation $n \to n\pm l$ in $\alpha_n^\pm$. When $\omega $ takes an integral value, we have the neutral string. We have seen how this occurs for the mode operastors. We have finally calculated the $g$-factors of charged strings in the framework of our formulation, which leads us to the  noncommutative strings. We have found $g_s =2$ for the string with any spin $s$, and $g_L=1$
for the same string with any orbital angular momentum $L$.

\newpage
\renewcommand{\theequation}{\Alph{section}.\arabic{equation}}
\setcounter{equation}{0}
\setcounter{section}{1}
\section*{Appendix A}
\indent

 In this Appendix we derive more explicit forms of functions $\bar{\epsilon}(\sigma:\gamma)$, 
 $E(\sigma, \sigma': \gamma)$ and also of $[b, b]$. \\
 \indent
 The function $\bar{\epsilon}(\sigma:\gamma)$ is defined by
\begin{equation}
\label{a1}
 \bar{\epsilon}(\sigma:\gamma)\equiv \frac{1}{\pi}\sum _{n}\frac{1}{\gamma-in}e^{(\gamma-in)\sigma}.
\end{equation}
Hence we have
\begin{equation}
\label{a2}
 \bar{\epsilon}'(\sigma:\gamma)=\frac{1}{\pi}\sum _{n}e^{(\gamma-in)\sigma}=2e^{\gamma\sigma}\delta(\sigma)
\end{equation}
and
\begin{eqnarray}
\label{a3}
 \bar{\epsilon}(0:\gamma)&=& \frac{1}{\pi}\lim_{N\to \infty}\sum _{n=-N}^N\frac{1}{\gamma-in}
 =\frac{i}{\pi}\lim_{N\to \infty}\sum _{n=-N}^N\frac{1}{n+i\gamma} \nonumber \\
 &=&i\cot (i\pi\gamma)=\coth\pi\gamma.
\end{eqnarray}
Define
\begin{eqnarray}
\label{a4}
 \hat{\epsilon }(\sigma:\gamma)\equiv 2\int _0^\sigma d\sigma'e^{\gamma\sigma'}\bar{\delta}(\sigma')
 =2\sum _{k=-\infty }^\infty \int _0^\sigma d\sigma' e^{2\pi k\gamma }
 \delta (\sigma'-2\pi k\gamma ) \nonumber \\
 =\sum _{k=-\infty }^\infty e^{2\pi k\gamma }\{\epsilon (\sigma-2k\pi)
 +\epsilon (2k\pi)\},
 \end{eqnarray}
where
\begin{equation}
 \epsilon (\sigma)\equiv \textrm{sgn}(\sigma). 
\end{equation}
Then we get
\begin{equation}
\label{a5}
 \bar{\epsilon}(\sigma:\gamma)=\hat{\epsilon}(\sigma:\gamma)+\coth\pi\gamma . 
\end{equation}
For $2n\pi<\sigma <2(n+1)\pi$, the first term in (\ref{a4}) can be written as
\begin{eqnarray*}
\sum _{k=-N}^{N}e^{2\pi k\gamma }\epsilon (\sigma-2k\pi)=\sum _{k=-N}^{n}e^{2\pi k\gamma }
 -\sum _{k=n+1}^{N}e^{2\pi k\gamma } \\
 =\frac{1}{e^{2\pi\gamma }-1}\{e^{2(n+1)\pi\gamma }-e^{-2N\pi\gamma }-e^{2(N+1)\pi\gamma }\}.
\end{eqnarray*}
On the other hand, the second term is
\begin{eqnarray*}
 \sum _{k=-N}^{N}\epsilon (2k\pi)e^{2\pi k\gamma }=\sum _{k=1}^{N}(e^{2\pi k\gamma }-e^{-2\pi k\gamma })
 =\frac{e^{2(N+1)\pi\gamma }-e^{2\pi\gamma }+e^{-2N\pi\gamma }-1}{e^{2\pi\gamma }-1}.
\end{eqnarray*}
Hence, we have
\begin{eqnarray}
\label{a6}
\hat{\epsilon }(\sigma:\gamma)&=&\frac{1}{e^{2\pi\gamma }-1}\{2e^{2(n+1)\pi\gamma }-e^{2\pi\gamma }-1\} \nonumber \\
 &=& \frac{e^{(2n+1)\pi\gamma }}{\sinh\pi\gamma }-\coth\pi\gamma \qquad (2n\pi<\sigma<2(n+1)\pi)
\end{eqnarray}
to give
\begin{equation}
\label{a7}
 \bar{\epsilon }(\sigma:\gamma)=\frac{e^{(2n+1)\pi\gamma }}{\sinh\pi\gamma }. \qquad (2n\pi<\sigma<2(n+1)\pi)
\end{equation}
At $\sigma=2n$ this is defined by a mean value of both sides
\begin{equation}
\label{a8}
 \bar{\epsilon }(2n\pi:\gamma)=\frac{e^{(2n+1)\pi\gamma }+e^{(2n-1)\pi\gamma }}{2\sinh\pi\gamma }
 =e^{2n\pi\gamma }\coth\pi\gamma  .
\end{equation}
From (\ref{a7}) and (\ref{a8}) the explicit formula of $\epsilon(\sigma:\gamma)$ is
\begin{equation}
\label{a9}
 \bar{\epsilon}(\sigma:\gamma)=\frac{e^{\pi\gamma }}{\sinh\pi\gamma }\sum_{k}e^{2\pi k\gamma }
 \{\theta (\sigma-2k\pi)-\theta (\sigma-2(k+1)\pi)\},
\end{equation}
where $\theta $ is the ordinary step function. \\
\indent
 The function $E(\sigma,\sigma':\gamma)$ is defined by (\ref{f11}), i.e.
\begin{eqnarray}
\label{a10}
 E(\sigma,\sigma':\gamma)\equiv \bar{\epsilon }(\sigma-\sigma':\gamma)
 +\bar{\epsilon }(-\sigma+\sigma':\gamma) \nonumber \\
 +e^{-2\beta_0}\bar{\epsilon }(\sigma+\sigma':\gamma)
 +e^{2\beta_0}\bar{\epsilon }(-\sigma-\sigma':\gamma).
\end{eqnarray}
For $0 < \sigma,\sigma' < \pi$, this becomes
\begin{equation}
\label{a11}
 E(\sigma,\sigma':\gamma)=\frac{2}{\sinh\pi\gamma }\{\cosh\pi\gamma +\cosh(\pi\gamma -2\beta_0)\}
 =\frac{4\cosh\beta_0\cosh\beta_\pi}{\sinh\pi\gamma }.
\end{equation}
In order that $X$ is commutable in this region, the constant term in (\ref{f22}) should be
\begin{equation}
\label{a12}
 i2c=\frac{1}{\pi}[b, b]=i\frac{\cosh\beta_0\cosh\beta_\pi}{\sinh\pi\gamma }.
\end{equation}
This coincides with Eq.(\ref{f23}).

\setcounter{equation}{0}
\setcounter{section}{2}
\section*{Appendix B}
\indent

 We summarize a consequence of the coefficient 1/4 in our Lagrangian.(2.4). For brevity we set 
 $B = 0$.  The Lagrangian is given by
\begin{equation}
\label{ab1}
 L=\frac{1}{4}\int _0^\pi d\sigma (\dot{X}^2-X'^2).
\end{equation}
The canonical momentum is defined by
\begin{equation}
\label{ab2}
P_c\equiv \frac{\partial L}{\partial \dot{X}}=\frac{1}{2}\dot{X}.
\end{equation}
The Poisson bracket and the Dirac bracket are then given by
\begin{equation}
\label{ab3}
(X(\sigma),P_c(\sigma'))_{PB}=\delta(\sigma-\sigma')
\end{equation}
and
\begin{equation}
\label{ab4}
(X(\sigma),P_c(\sigma'))_{DB}=\frac{1}{2}\delta_c(\sigma, \sigma')
\end{equation}
respectively. (Eq.(B.4) is obtained by taking $B \to 0$ in Eq.(6.19).) \\
\indent
    We define the Virasoro operator as
\begin{equation}
\label{ab5}
 \tilde{L}_n=\frac{1}{8}\int _{-\pi}^{\pi}d\sigma e^{\pm in\sigma}(\dot{X} \pm X')^2
 =\frac{1}{2}\int _{-\pi}^{\pi}d\sigma e^{\pm in\sigma}(P_c \pm \frac{1}{2}X')^2,
\end{equation}
which leads us to
\begin{equation}
\label{ab6}
 (X(\sigma),\tilde{L}_n)_{PB}=\dot{X}(\sigma)\cos n\sigma +iX'(\sigma)\sin n\sigma
\end{equation}
for the Poisson bracket, and to
\begin{equation}
\label{ab7}
 (X(\sigma),\tilde{L}_n)_{DB}=\frac{1}{2}(\dot{X}(\sigma)\cos n\sigma +iX'(\sigma)\sin n\sigma)
\end{equation}
for the Dirac bracket. In view of 1/2 in (B.7) one can see that the conformal generator in the Dirac quantization should be 
\begin{equation}
\label{ab8}
 L_n=2\tilde{L}_n=\frac{1}{4}\int _{-\pi}^\pi d\sigma e^{\pm in\sigma }:(\dot{X}\pm X')^2:.
\end{equation}
\indent
 To sum up, it is convenient to use the modified momentum
\begin{equation}
\label{ab9}
 P\equiv 2P_c =\dot{X}
\end{equation}
to give
\begin{equation}
\label{ab10}
(X(\sigma),P(\sigma'))_{DB}=\delta _c(\sigma,\sigma'),
\end{equation}
\begin{equation}
\label{ab11}
 L_n=\frac{1}{4}\int _{-\pi}^{\pi}d\sigma e^{\pm in\sigma}(\dot{X} \pm X')^2
 =\frac{1}{4}\int _{-\pi}^{\pi}d\sigma e^{\pm in\sigma}(P \pm X')^2.
\end{equation}
The results coincide with those in the conventional quantization.


\end{document}